
\def\frac#1#2{{#1\over#2}}

\magnification=\magstep1   \openup1\jot  \vsize=9truein
\font\titlefont=cmcsc10 scaled \magstep2

\centerline {\titlefont The Superscattering Matrix}\centerline{
\titlefont For Two Dimensional Black Holes }  \vskip 1truein
\centerline {S.~W.~Hawking} \vskip .5truein  \centerline
{Department of Applied Mathematics and Theoretical Physics}
\centerline {University of Cambridge}  \centerline {Silver
Street}  \centerline {Cambridge CB3 9EW}  \centerline {UK}

\vskip .25truein
\centerline{California Institute of Technology}
\centerline{Pasadena CA 91125}
\centerline{USA}

\vskip .5truein  \centerline {\it November 1993 }  \vskip
1truein  \centerline {\bf Abstract}
 \bigskip
  A consistent Euclidean semi classical calculation is given for
the superscattering operator $\$ $ in the RST model for states
with a constant flux of energy. The $\$ $ operator is CPT
invariant. There is no loss of quantum coherence when the energy
flux is less than a critical rate and complete loss when the
energy flux is critical.
 \vfill \eject
  \beginsection 1. Introduction

In classical general relativity a lot of information is lost in
the formation of a black hole. A black hole will settle down
rapidly to a stationary state. The classical no hair theorems
say that this state is characterized just by the values of
conserved charges, like mass, angular momentum, and electric
charge, that are coupled to gauge fields. In other words,  a
black hole of a given mass, angular   momentum, and electric
charge, can be formed by the collapse of a very large number of
different objects.

This loss of information was not a worry in the purely classical
theory, because one could say the information was still inside
the black hole, even if one couldn't get at it. But the
situation changed when it was discovered  that according to
quantum theory, black holes should radiate and slowly evaporate.
This made the question about the information in a black hole
much more pressing. If the black   hole evaporated and
disappeared completely, what would happen to the information?
There were three possible ways the information might be
preserved:

\item 1 The information might come out again at the end of the
evaporation. The problem was that information requires energy to
carry it  and there wouldn't be much energy left in the final
stages.

\item 2 The information might come out continuously during the
evaporation. The difficulty here was that the information would
be carried by the in-falling matter, far beyond the apparent
horizon. So, if it were to appear outside the horizon, it would
violate causality. If one could send information faster than
light like this, one could also send it back in time.

\item 3 The black hole might not evaporate completely, but might
leave some long lived remnant that could be said still to
contain the information. But this would violate CPT, if black
holes could form, but never disappear completely.

In this paper I want to take seriously the idea that information
is lost. As I showed some time ago[1], this would imply a loss
of quantum coherence. One could form a black hole from matter in
a pure quantum state, but it would decay into radiation in a
mixed quantum state described by a density matrix. In ordinary
quantum field theory the evolution is described by an $S$
matrix. This can be thought   of as a two index tensor on the
Hilbert space that maps initial states to final states: $$ \psi
\sb f {}^A= S^A{}_B\psi _i{}^B$$  However when one goes to
quantum gravity, the possibility of forming real or virtual
black holes means that the evolution is given by what I call a
super scattering operator, $\$$. This can be thought of as a
four index tensor on Hilbert space that maps initial density
matrices   to final ones: $$ \rho \sb f {}^A{}_B=
\$^A{}_{BC}{}^D\rho _i{}^C{}_D$$ In general, the $\$$ operator
will not be the product of an $S$ matrix with its complex
conjugate: $$ S^A{}_{BC}{}^D\ne S^A{}_C\bar S_B{}^D$$ This
proposal of a non unitary evolution was greeted with outrage by
most particle physicists. I was accused of violating quantum
mechanics. That is not the case.  One gets loss of quantum
coherence and a mixed state whenever there is part of a system
you don't measure. All I have done, is point out you can't
measure the part of the quantum state that is inside the black
hole.

The argument about whether quantum coherence is lost in black
hole evaporation has dragged on inconclusively over the years
because general relativity is not renormalizable. This has meant
we have not been able to calculate what happens in the final
stages of black hole evaporation. Maybe the answer lies in
supergravity or superstring theory, but we don't know how to use
them to do the calculation.   However, in the last two years,
there has been a revival of interest in two dimensional black
holes. These have the great advantage of being renormalizable,
so it should be possible to use them to decide the issue. Those
who felt a mission to defend quantum purity, hoped that these
two dimensional models would show that information and quantum
coherence  were preserved. However, they have been
disappointed   with the results so far: all calculations to date
have shown loss of information and quantum coherence.  This has
left the unitary $S$ matrix camp mumbling something about the
breakdown of the large N approximation, but with no real
argument.

Most of the calculations that have been done on two dimensional
black holes, have been carried out in Lorentzian spacetime. They
have assumed no horizons or singularities in the past. One can
show[2] that this implies that there must be horizons or naked
singularities in the future. Thus the calculations are
manifestly not CPT invariant. This is reflected in the fact that
the outgoing energy flux they   predict is always below a
certain critical level. However, the ingoing energy flux can
have any value. Thus the super scattering operator given by
these calculations will not be CPT invariant. To lose quantum
coherence is bad enough, but to lose CPT as well, sounds like
carelessness.

I shall therefore outline a different approach to calculating
the super scattering operator which is guaranteed to give CPT
invariant results. It is the Euclidean path integral method,
coupled with the no boundary condition[3]. In the cosmological
case, the no boundary condition literally meant that spacetime
had no boundary. In other words, the quantum state is defined by
a path integral over all   compact positive definite metrics.
But in the particle scattering case, the appropiate quantum
state is defined by a path integral over all positive definite
metrics that have one of more asymptotically Euclidean regions
but no other boundary. One can show that the asymptotic Green
functions in this quantum state are CPT invariant[4]. It then
follows that the super scattering operator is CPT invariant,
in the sense of detailed balance[5]:  the probability to go from
the initial pure state A to a final state B, is the same as the
probability to go from the CPT conjugate of B to the CPT
conjugate of A.

\beginsection 2. Two dimensional black holes

The starting point for two dimensional black holes is the CGHS
model[6]. This has a metric, $g$, and dilaton field, $\phi$,
coupled to $N$ minimal scalar fields, $f_i$. $$L={1\over
2\pi}\sqrt{-g}\left[e^{-2\phi}(R+4(\nabla\phi
)^2+4\lambda^2)-{1\over 2}\sum_{i=1}^N(\nabla f_i)^2\right]$$
The quantum effective action of the minimal scalars in the
curved metric, $g$, can be evaluated exactly and added to the
action of the classical CGHS model. One can then define new
field variables, $\chi$ and $\Omega$, in   which the theory is a
conformally invariant quantum field theory and the action looks
rather like that of the Liouville model: $$S={1\over\pi}\int
d^2x\left[-\partial_+\chi\partial_-\chi
+\partial_+\Omega\partial_-\Omega
+\lambda^2\exp\left[{2\over\sqrt\kappa}(\chi -\Omega
)\right]+{1\over
2}\sum_{i=1}^N\partial_+f_i\partial_-f_i\right]$$  In making the
theory, a conformal field theory, one has to modify the action.
One can either change the kinetic term, as Russo, Susskind, and
Thorlacius do[7], or the non derivative term, as Bilal and
Callan[8], and de Alwis[9] do. Both modifications end up with
the same Liouville like theory, but the relation between the
Liouville fields, $\chi $ and $ \Omega $, and the physical
fields, $g$ and $\phi $, is different. The RST version has the
great advantage that it admits the linear dilaton as a solution.
The other, ABC, version does not seem to have any natural ground
state. I shall therefore work with the RST version.

In the RST model the Liouville fields are related to the physical
fields by $$\Omega ={\sqrt\kappa\over 2}\phi
+{e^{-2\phi}\over\sqrt\kappa}$$ $$\chi = \sqrt\kappa\rho
-{\sqrt\kappa\over 2}\phi +{e^{-2\phi}\over\sqrt\kappa}$$ where
the metric is in the conformal gauge $$ds^2=-e^{2\rho}dx^+dx^-$$
The Liouville model looks to be almost linear, but this is
deceptive. Both in the ABC and RST versions, the field, $ \Omega
$, is bounded below as a function of the physical field $\phi $.
It takes its minimum value, $\Omega _c $, when $\phi = \phi _c
$. There are two possible attitudes one can take to this. One,
propounded by de Alwis[9], is that one should take the quantum
theory to be defined by a path integral over the full range of
$\chi $ and $ \Omega $ from minus infinity to plus infinity. The
other attitude, put forward by RST[7],  is that the path
integral should be restricted to the range of $\Omega $ that
corresponds to real physical   fields. I shall adopt this latter
approach. The physical fields are what are important. The
Liouville fields, $\chi $ and $ \Omega $, can be regarded just
as mathematical constructions that help to solve the field
equations, but are of no physical significance of themselves.

The field equations for the Liouville model are simple.
$$\partial _+ \partial _- ( \chi - \Omega)=0$$ $$\partial _+
\partial _- ( \chi + \Omega)=-{2\lambda ^2\over \sqrt
\kappa}\exp \left({2\over \sqrt \kappa}(\chi - \Omega)\right)$$
The first equation corresponds to the freedom to choose
coordinates. One can therefore take $$ \chi - \Omega =0$$ at
least locally. It is then easy to write down the   general
solution in this coordinate gauge: $$ \Omega = -{ \lambda
^2\over \sqrt \kappa} x_+x_- + F(x_+)+G(x_-)$$ To make this
solution Euclidean, introduce a complex coordinate $z $ in the
plane with $$z=x_+, ~~~~\bar z =-x_-$$

This very simple theory becomes highly non trivial if one
imposes the requirement that the physical fields be non
singular. One way to get a non singular solution is if omega is
always above its minimum value, $\Omega _c$.  The only solutions
like this that are also static and with only an asymptotically
flat region can be written down in the $\chi = \Omega $ gauge:
$$ \Omega ={\lambda ^2\over \sqrt   \kappa}z\bar z +M +{\sqrt
\kappa \over 2} $$ $\Omega $ is a function only of $r$  in polar
coordinates, with a minimum at the origin. These solutions
represent Euclidean black holes with the horizon at the origin.
The polar angle, $\theta $ corresponds to imaginary time.
Because this is periodic, the black hole is in thermal
equilibrium. There is a steady flow of energy in at infinity and
a similar flow out.

Suppose now that one has a field configuration in which $\Omega $
actually reaches the critical value on some curve, $C$.  Then
$\Omega $ will drop below the critical value in a neighbourhood
of $C$ and the physical fields will be singular unless the
gradient of $\Omega $  vanishes on $C$. Thus the necessary
condition for the physical fields to be non singular is
$$\left.\nabla \Omega \right|_{C}=0$$   I shall refer to this as
the RST boundary condition. With this boundary condition, no
signal can propagate through the curve $C$. One can therefore
cut off the spacetime beyond $C$ and consider only the region
between $C$ and asymptotically flat infinity. One can regard the
curve $C$ as being like the axis of symmetry in the dimensional
reduction of a spherically symmetric spacetime to two
dimensions.   On the basis of this analogy, the appropiate
boundary condition on the minimal scalars would be that their
normal gradient should vanish on $C$.

If $C $ is time like, the normal direction will be space like.
The field equation  $$\partial _+ \partial _- \Omega=-{\lambda
^2\over \sqrt \kappa}\exp \left({2\over \sqrt \kappa}(\chi -
\Omega)\right)$$ will then imply that the second derivative of
$\Omega $ in the normal direction will be positive. This means
that $\Omega \ge \Omega _c$ in a neighbourhood of $C$ and the
physical fields can be non singular. On the other hand, if $C $
is space like, the normal direction will be time like and the
second derivative of $\Omega $ in the normal direction will be
negative. Thus $\Omega $ will go below $ \Omega _c$ and the
physical fields will be singular. Thus the RST boundary
condition can not be imposed in Lorentzian spacetime if the
boundary is space like.

By contrast, in the Euclidean regime the normal direction to $C$
is always space like. This means  one can always satisfy the
boundary condition. Thus the Liouville model with the RST
condition and the no boundary condition, is a well defined
quantum theory. One should therefore be able to calculate the
superscattering operator and see whether it involves loss of
quantum coherence. I shall give evidence   that it does, in at
least some situations.

\beginsection 3. The quantum RST model

If one just had the Liouville model without any restriction on
the range of $\Omega $, the theory would be linear and the semi
classical approximation would be exact. This means that the
quantum theory would be determined completely by solutions of
the quantum effective action, which will be the same as the
original Liouville action. However, if the range of $\Omega $ is
restricted, and the RST boundary   condition imposed, the theory
becomes effectively non linear. This means that the solutions of
the Liouville theory, are no longer the whole story. But one can
still hope that they will give a first approximation.

As well as the field equations obtained by varying $\phi $ and
$\rho $,  there are constraint equations obtained by varying the
components of the metric that are zero in the conformal gauge.
In terms of the physical fields, the constraint equations can be
written in a form like the Einstein equations:
$$-\left[e^{-2\phi}+{\kappa\over
4}\right](4\partial_\pm\rho\partial_\pm\phi-2\partial^2_\pm\phi
)={1\over 2}\sum_{i=1}^N \partial_\pm f_i\partial_\pm
f_i-\kappa(\partial_\pm\rho\partial_\pm\rho
-\partial_\pm^2+t_\pm )$$ The term on the left that is the
analogue of the Einstein tensor, is the trace free part   of the
second covariant derivative of $\phi $.  The first term on the
right is the classical energy momentum tensor of the scalar
fields. The remaining terms on the right can be regarded as the
quantum induced energy momentum tensor. It is non local because
it contains the functions, $t_{\pm}(x_{\pm})$, that have to be
determined by boundary conditions. Given a solution of the field
equations, one   can always find functions $t_{\pm}(x_{\pm})$
that will satisfy the constraint equations. I shall therefore
regard the left hand side of the constraint equations as the
{\it definition} of the total energy momentum tensor, classical
plus quantum, of the scalar fields. In other words, given a
solution of the $\phi $ and $\rho $ field equations, one can
read off from the constraint equations what the   energy
momentum flow is.

In the $\chi = \Omega $ gauge there are a two parameter family of
solutions which have a Killing vector in the direction of the
polar angle, $\theta $. The two parameters are the coefficient
of the $\log r $ term  and the constant term in $\Omega $.
Imposing the RST boundary condition that $\nabla \Omega =0$
where $\Omega = \Omega _c$ puts one relation between the two
parameters. One therefore has   a one parameter family of static
Euclidean solutions. $$\Omega ={\lambda ^2\over\sqrt\kappa}z\bar
z -P\log (\lambda ^2 z\bar z
)-P\left[\log\left(P\sqrt\kappa\right)+1\right]-{\sqrt\kappa\over
4}\left[\log\left({\kappa\over 4}\right)+1\right]$$ The linear
dilaton is the  member of this family with
$P=\frac{\sqrt\kappa}{4}$.  In it, $\Omega $  has a minimum of
$\Omega _c $ on a circle around the origin. The total energy
momentum tensor of the scalar fields calculated in the manner I
described, is zero. Thus this solution represents the vacuum
with no energy flux at infinity. However for $0\le
P<\frac{\sqrt\kappa}{4}$, there is a family   of solutions with
a constant positive flux of energy coming in at infinity and a
similar flux going out. As the energy flux increases from zero,
the circle at which $\Omega = \Omega _c $ shrinks in radius.
When the flux reaches the critical level, $ $ the circle shrinks
to zero and the solution becomes a Euclidean black hole with no
boundary and with $\Omega > \Omega _c $ everywhere. There is a
one parameter   family of black hole solutions, but all have the
same energy flux at infinity. There are no regular static
Euclidean solutions with an energy flux greater than this value.
That is what one might expect because if one sent in energy at
such a rate, one would create a black hole that would grow
faster than it could radiate itself away. So one couldn't get a
static solution.

\beginsection 4. The superscattering operator

One would expect these Euclidean solutions to give the dominant
contributions to the superscattering operator for incoming
states with a steady flux of energy.  Consider first the
solutions, like the linear dilaton, that have a boundary at
$\Omega _c$ In these, the region inside the boundary circle is
not regarded as part of the spacetime. This means there's no
reason to identify the polar angle, $\theta   $, with any
particular period. It is therefore natural to take $\theta $  to
run the full range from minus infinity to infinity.

It is straightforward to calculate Green functions for the scalar
fields in these spacetimes. The minimal scalars propagate in the
flat background metric independently of the dilaton and the
conformal factor. They are however affected by the global
structure such as whether $\theta $ is periodically identified
and by the presence of the boundary. On the basis that the
boundary is like an axis of symmetry,   it is natural to impose
Neumann boundary conditions on the minimal scalars: the normal
gradient of each field vanishes on the boundary. It is then
straightforward to obtain the Green functions by the method of
images.

The procedure to calculate the superscattering operator from the
asymptotic Green functions was described in [5]. One first
defines creation and annihilation operators for incoming and
outgoing particles in terms of integrals of the field operators
at infinity, e.g. $$a_i(\omega )=-{i\over \sqrt{2\pi}}
\int_{\Sigma}p_i(\omega,
x){\buildrel\leftrightarrow\over\nabla}_{\mu} f(x) d \Sigma
^{\mu}$$   where $a_i(\omega)$ is the annihilation operator for
a particle in the incoming mode $p_i(\omega, x)$ of frequency
$\omega $, $\Sigma $ is a time like line near infinity and the
integral is carried out over real Lorentzian time. The
superscattering operator is then the expectation value of a
string of creation and annihilation operators for the initial
and final states and their complex conjugates: $$
\$^A{}_{BC}{}^D=\langle O_C O^A O_B O^D\rangle $$ where $O^D $
is a string of creation operators for the   initial state $|
\psi _i{}^D\rangle =O^D| 0\rangle $ and $O_C$ is a string of
annihilation operators for the initial state $\langle \psi
_i{}_C| =\langle 0| O_C$. Similarly $ O^A$ and $O_B$ are the
creation and annihilation operators for the final state.

One interprets the expectation values $\langle f(x_1),f(x_2),...
f(x_n)\rangle $ as the analytic continuation of the Euclidean
Green function $G(x_1,x_2,...  x_n)$ keeping each of the
differences $x_{i+1}-x_i$ as a vector with a small future
directed imaginary part. In other words, to evaluate the
superscattering matrix element $\$^A{}_{BC}{}^D$ one does a
multiple integral of the analytically continued Green   function
on the time like line $\Sigma $ near infinity with the contours
of integration arranged in increasing order of imaginary time
reading from left to right in the expectation value.

 In ordinary non gravitational quantum field theory, the
superscattering operator factorizes  $$ \$ ^A{}_{BC}{}^D=\langle
O_C O^A \rangle \langle O_B O^D\rangle $$ The second factor on
the right is defined to be the $S$ matrix and the first factor
is its complex conjugate. There is no loss of quantum coherence
and the $S$ matrix is unitary. But in quantum gravity, the
superscattering matrix may not   factorize leading to loss of
quantum coherence, as I shall show.

The static solutions with a boundary are conformal to half of
flat space. This means that the Green functions for the scalar
fields are the same as those for flat space with a reflecting
wall. They will propagate a positive frequency wave function
purely in the negative imaginary time direction, and a negative
frequency wave function purely in the positive imaginary time
direction. This means that   there will be propagation only
between the top two contours corresponding to $O^D$ and $O_B$
and between the bottom two contours corresponding  to $O^A$ and
$O_C$. Thus the superscattering operator will factorize and
there will be no loss of quantum coherence. One can adjust the
solution so that the ingoing flux of the initial state is equal
to the energy flux of the solution, calculated in the way   I
described. This gives a consistent semi classical calculation of
the superscattering operator for initial states with constant
energy fluxes less than the critical flux.

The situation is very different however, if one considers Green
functions on a black hole background. In this case, the point at
the origin of Euclidean space is part of the physical spacetime.
In order for the metric to be regular there, the polar angle,
$\theta $ must be identified with a period of $2\pi $.  This
means that the asymptotic Green functions will be periodic in
imaginary time, with a   period $\beta ={2\pi\over\lambda}$. The
propagation of positive frequencies will still be in the
negative imaginary time direction. But now there will be images
of the contours of integration, periodically spaced in the
imaginary time direction. This means that the propagation need
not be just from the $O^D$ contour to the $O_B$ one and from the
$O^A$ contour to the $O_C$ one. Instead there can be a   non
zero amplitude for the initial creation operators $O^D$ to
propagate to the initial annihilation operators $O_C$ and for
the final creation and annihilation operators to do is same.
This means that the superscattering operator will not factorize,
so there will be loss of quantum coherence.

The energy momentum tensor for the scalar fields is $$T_{\pm
\pm} =\sum \partial _{\pm} f\sb i\partial _{\pm}f\sb i-{N \over
12}\left(\partial _{\pm} \rho \partial _{\pm} \rho - \partial
^2_{\pm \pm}+t_{\pm}(x_{\pm})\right)$$ The functions
$t_{\pm}(x_{\pm})$ should be chosen to make the second term on
the right zero asymptotically. This will mean that the energy
flux at infinity is just given by first   term on the right,
which is the classical energy momentum tensor of the $f_i$
fields, which are given by propagating the data for the initial
and final states with the analytically continued Euclidean
propagator. The total energy momentum tensor will be regular on
the Euclidean space except at the origin where it will in
general have a singularity. This singularity will however vanish
if the classical   energy momentum tensors of the initial and
final states both have exactly the critical energy flux. Thus in
this case, and only in this case, one will have a consistent
semi classical calculation of the superscattering operator,  but
this time with loss of quantum coherence. For suppose that the
initial state was a pure quantum state in one particular scalar
field. Then the final state could be any   state, in any
combination of the scalar fields, that has the critical energy
flux. But this is just the condition for a thermal spectrum. So
the super scattering operator takes any initial pure state with
the critical energy flux to a final mixed thermal quantum state.
Thus there is loss of quantum coherence.

These  semi classical calculations give no loss of quantum
coherence,  when the energy flux is below the critical value,
and complete loss, when it is critical. But one would expect
that a full calculation, would give a gradual transition. As the
energy flux tends towards the critical value, the boundary at
$\Omega \sb c$ in the semi classical solutions, becomes a curve
of greater and greater acceleration.   It would require only a
small quantum fluctuation in $\Omega $ to make the transition to
a black hole metric with no boundary. Thus one would expect a
gradually increasing amount of quantum incoherence, as the
energy flux approached the critical value.

In this paper I have been considering the superscattering
operator for steady fluxes of energy. But what one really wants
to consider are bursts with finite total energy, but with a
maximum flux above the critical level. It seems that a
Lorentzian semi classical solution for this situation will have
a boundary at $\Omega \sb c$ that is space like. Thus the RST
boundary condition can not be satisfied. This   means that the
theory is not well defined: in the absence of a boundary
condition on part of the boundary, one can not calculate Green
functions, and predict the results of scattering. This lack of
predictability, is usually minimized by assuming the metric is
non singular in the past. But then there is necessarily a
singularity in the future, at least one point of which must be
naked.  This is rather   ad hoc, and unsatisfactory. It is also
clearly not CPT invariant. To restore CPT symmetry, one should
presumably consider metrics with singularities in both the past
and future[10]. But then the lack of a boundary condition at the
past singularity, which would be naked, would mean that one was
completely unable to predict the final state. The only way out,
would seem to be to impose boundary conditions   in the
Euclidean regime. There will still be singularities in the
Lorentzian regime, but now the behavior of the Green functions
at the singularities will be determined by analytic continuation
from the Euclidean regime. Hopefully, this will give physically
reasonable, CPT invariant results, for the scattering of bursts,
probably with loss of quantum coherence. Further work on this is
in progress.

I'm very grateful for discussions with James Grant, Justin
Hayward, Andy Strominger and Edward Teo. This work was partly
carried out at Caltech where I was a Sherman Fairchild
distinguished scholar.

\beginsection References

\par

 1.  Hawking, S.W., {\it Breakdown of Predictability in
Gravitational Collapse}, Phys.Rev. D{\bf 14}, 2460 (1976).

 2.  Hawking, S.W. and Hayward, J., {\it Quantum Coherence in Two
Dimensions}, preprint DAMTP-R93/15 (hepth 9305165).

 3.  Hawking, S.W. and Hartle, J.B., {\it Wave Function of the
Universe}, Phys.Rev.D{\bf 28}, 2960 (1983).

 4.  Hawking, S.W., {\it The Unpredictability of Quantum
Gravity}, Commun.Math.Phys.{\bf 87}, 395 (1982).

 5.  Wald, R.M., {\it Black Holes, Thermodynamics and Time
Reversibility}, Quantum Gravity 2, Clarendon Press, Oxford, 223
(1981).

 6.  Callan, C.G., Giddings, S.B., Harvey, J.A. and Strominger,
A., {\it Evanescent Black Holes}, Phys.Rev.D{\bf 45}, 1005
(1992).

 7.  Russo, J.G., Susskind, L. and Thorlacius L., {\it End Point
of Hawking Radiation}, Phys.Rev.D{\bf 46}, 3444 (1992).

 8.  Bilal, A. and Callan, C.G., {\it Liouville Models of Black
Hole Evaporation}, Nucl. Phys. B {\bf 394}, 73 (1993).

 9.  De Alwis, S.P., {\it Quantum Black Holes in Two Dimensions},
Phys.Rev.D{\bf 46}, 5429 (1992).

10.  Stominger, A., {\it White Holes, Black Holes and CPT in Two
Dimensions}, preprint NSF-ITP-93-92 (hepth 9307079).

\end